\documentclass[pra,twocolumn,showpacs,amssymb]{revtex4}
\usepackage[dvipdfm]{graphicx}
\usepackage{amsmath}
\usepackage{pdfpages, epsfig}

\begin{document}
\title{Floquet Edge States with Ultracold Atoms}

\author{Matthew D. \surname{Reichl}}
	%\affiliation{Laboratory of Atomic and Solid State Physics, Cornell University, Ithaca, New York 14853, USA}
	
\author{Erich J. \surname{Mueller}}
	\affiliation{Laboratory of Atomic and Solid State Physics, Cornell University, Ithaca, New York 14853, USA}

\date{\today}


\begin{abstract}        % give a summary of your paper
 We describe an experimental setup for imaging topologically protected Floquet edge states using ultracold bosons in an optical lattice. Our setup involves a deep two dimensional optical lattice with a time dependent superlattice that modulates the hopping between neighboring sites. The finite waist of the superlattice beam yields regions with different topological numbers. One can observe chiral edge states by imaging the real-space density of a bosonic packet launched from the boundary between two topologically distinct regions.
%                         please supply keywords within your abstract
\end{abstract}
\maketitle

\section{Introduction}

One of the most exciting prospects in ultracold atomic physics is the ability to experimentally engineer and probe quantum states with topological order. Many of the theoretical proposals in this direction involve synthetic gauge fields \cite{lin2009, dalibard2011} which require complicated experimental setups in which Raman lasers couple internal atomic degrees of freedom. More recently there have been proposals to generate topological order in Floquet systems with periodically driven optical lattices \cite{hauke2012,  koghee2012,  zhang2014, zheng2014, baur2014, creffield2014}. While every technique brings its own technical challenges, the Floquet approaches appear to be simpler. Similar proposals appear in the solid state and photonics literature \cite{lindner2011, tong2013, gomez2013, ho2014}. Recent cold atoms experiments have successfully demonstrated uniform 1D gauge fields \cite{struck2012} and band hybridization \cite{parker2013} using shaken optical lattices. Time-periodic Hamiltonians with effective magnetic fields have also been implemented with Raman techniques \cite{aidelsburger2013, miyake2013}.

In this paper we propose an experiment that simulates an especially simple square-lattice Floquet Hamiltonian \cite{rudner2013} that nonetheless displays edge state physics and topological order. We discuss an implementation using bosons in an optical lattice, and demonstrate using numerical simulations how edge states can be directly imaged in the system.

Our proposal for probing edge states in this system involves initializing and releasing a wave packet of bosons at the spatial boundary between two topological phases and directly observing chiral edge states by watching how this packet evolves. This proposal can be thought of as the cold atom analogue to a recent quantum optics experiment \cite{rechtsman2013} where wave packets of light were directly observed propagating along the edge of a topological Floquet system. Here bosonic atoms play the role of the photons. Our imaging scheme is similar to the proposal in Ref.~\cite{goldman2013} where propagating edge states are also directly observed in the density of atoms following the initialization and release of a wave packet. In that study, the authors considered non-driven topological systems. Our work extends this basic imaging idea to Floquet topological insulators. There are also connections between this approach and ideas of directly measuring Chern numbers by following wave-packet dynamics \cite{price2012, killi2012, dauphin2013, atala2013, wang2013}.

\section{Model} \label{secmod}

\begin{figure} \vspace{1.0em}
\hbox{\hspace{-0.55em}
\includegraphics[width=0.5\textwidth]{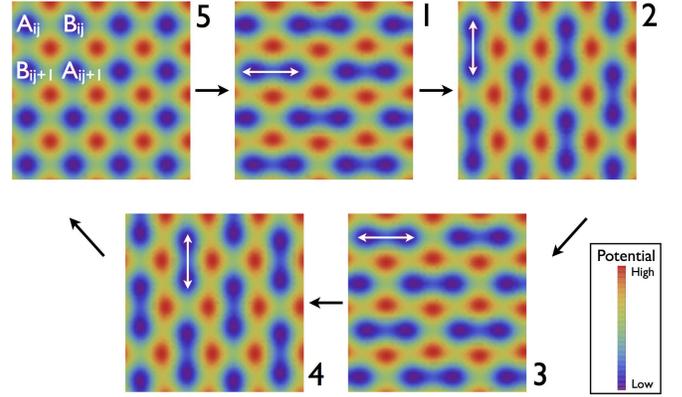}}
\caption{(Color online) Snapshots of the potentials used to produce the tight binding model for trapped atoms studied in this paper (Eq.~\ref{meq}).  Each potential is applied sequentially for a fixed period of time. As shown in the key, blue and red respectively correspond to low and high potential. As depicted by the white arrows, during time-steps 1, 2, 3, and 4, hopping only occurs between closely spaced ``dimers". No hopping occurs during time-step 5. The labeling of sites $A_{ij}$ and $B_{ij}$ are illustrated in panel 5. For the protocol described here, step 5 plays no role, but is convenient for generalizations that include a potential bias between A sites and B sites \cite{rudner2013}.} 
\label{modelimg}
\end{figure}

\begin{figure*} [t]
\centering
\includegraphics[width=1.0\textwidth]{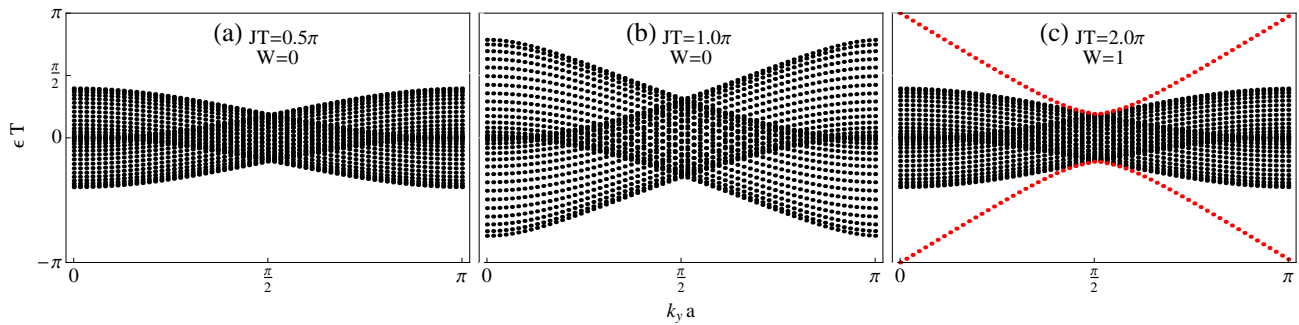}
\caption{(Color online) Floquet band structure of the model in Eq.~\ref{meq} with open boundary conditions in the $x$-direction and periodic boundary conditions in the $y$-direction. For weak hopping ($JT < J_c T= \frac{5}{4} \pi$: panels (a) and (b)) there are two overlapping bulk bands and the winding number $W=0$ (see Eq.~\ref{windeq}). For larger $J$, two edge modes are apparent (shown in red), and $W=1$. [cf. Ref.~\cite{rudner2013}].} 
\label{bandimg}
\end{figure*}

The simplest description of our proposed experiment is in terms of a two-dimensional tight binding model given by a Hamiltonian $H(t)$

\begin{equation} \label{meq}
\begin{split}
H(t) =\sum_{ij} &J_1(t) [c_{B_{i,2j-1}}^{\dagger} c_{A_{i+1,2j-1}} +c_{B_{i,2j}}^{\dagger} c_{A_{i,2j}} +h.c.] \\
                        &+ J_2(t) [c_{B_{i,j+1}}^{\dagger} c_{A_{i, j}} + h.c.] \\
                        &+ J_3(t) [c_{B_{i, 2j-1}}^{\dagger} c_{A_{i, 2j-1}} +c_{B_{i+1, 2j}}^{\dagger} c_{A_{i, 2j}} +h.c. ] \\
                        &+ J_4(t) [c_{B_{i,j}}^{\dagger} c_{A_{i, j+1}} + h.c.] \\
\end{split}
\end{equation}
where 

\begin{equation}
 J_{m}(t)= \begin{cases} J, & \mbox{if } \frac{(m-1)}{5} T<t\mod{T} < \frac{m}{5}T \\ 0, & \mbox{otherwise} \end{cases}
 \end{equation}
are the time dependent hopping parameters, and $T$ is the period of $H(t)$. Fig.~\ref{modelimg} gives a pictorial representation of this model. The fifth time interval, where $H(t)=0$, is unnecessary for our proposal but is included here in order to connect with prior literature \cite{rudner2013}. Section~\ref{secexp} describes how this model can be realized.

The sudden jumps between different hoppings are unnecessary, but make the analysis simpler. This model is readily generalized to the case where the hoppings $J_m(t)$ vary continuously with time. In an experiment the jumps can be quite sharp: For a tight binding model to be applicable, all time dependence need only be slow compared to the band spacing.

This same model is discussed in detail in Ref.~\cite{rudner2013}. Here we review some of its essential features. 

Because of the periodicity of the Hamiltonian $H(t+T)= H(t)$, we can use Floquet's theorem to express solutions of the time-dependent Schrodinger equation in the form $|\psi (t) \rangle =\exp(-i\epsilon t) |\phi(t) \rangle $ where $|\phi(t+T) \rangle= |\phi(t) \rangle$.  These states are eigenstates of the time evolution operator $U(t)$ evaluated at $t=T$: $U(T) |\psi(0)\rangle \equiv |\psi(T)\rangle = \exp(-i\epsilon T) |\psi(0)\rangle$. Often  $U(T)$ is calculated using perturbation theory in $T$, but here we find $U(T)$ exactly.

We can then define an effective time-independent ``Floquet Hamiltonian" $H_{\rm{eff}}= i \log(U(T))/T$. The branches of the log are chosen so that the eigenvalues of $H_{\rm{eff}}$, the quasi-energy spectrum $\epsilon$, fall within $-\pi/T < \epsilon \leq \pi/T$. One can take energy-space to be periodic, identifying $\epsilon$ with $\epsilon + \frac{2\pi n}{T}$ for any integer $n$. If we impose periodic boundary conditions on Eq.~\ref{meq}, $H(t)$ has the following form in momentum space:

\begin{equation} \label{kspaceeq} 
\begin{split}
& H(t)= \sum_{\mathbf{k}} (c_{\mathbf{k},A}^\dagger c_{\mathbf{k},B}^\dagger ) H(\mathbf{k},t) \begin{pmatrix} c_{\mathbf{k},A} \\ c_{\mathbf{k},B}  \end{pmatrix} \\
& H(\mathbf{k},t)= -J \sum^4_{n=1} (e^{i \mathbf{b}_n \cdot \mathbf{k}} \sigma^{+} + e^{-i  \mathbf{b}_n \cdot \mathbf{k}} \sigma^{-}) 
\end{split}
\end{equation}
where $\sigma^{\pm}= (\sigma_{x} \pm i \sigma_y)/2$, $\mathbf{b}_1= - \mathbf{b}_3 = (a,0)$, $\mathbf{b}_2= - \mathbf{b}_4 = (0,a)$, and $a$ is the nearest-neighbor lattice spacing. We can then label the quasi-energies $\epsilon(\mathbf{k})$,where $\mathbf{k}$ is the quasi-momentum. Of particular importance is the structure of $H_{\rm{eff}}$ near $\mathbf{k}=\mathbf{0}$ which is given by

\begin{eqnarray} \label{hsmallk}
H_{\rm{eff}}T  &\approx& \pi+ (\pi-{\textstyle \frac{4 JT}{5}}) \sigma_x  \\ \nonumber&&+
 f(J) \left[k_- \sin\left({\textstyle{\frac{JT}{5}}}\right) \sigma_y + k_+ {\textstyle \cos\left(\frac{JT}{5}\right)} \sigma_z\right]
%\begin{split}
%& (H_{\rm{eff}}T-\pi)  \approx  (\pi-\frac{4 JT}{5}) \sigma_x + \\
%& f(JT) [a(k_x-k_y) \sin(\frac{JT}{5}) \sigma_y + a(k_x+k_y)  \cos(\frac{JT}{5}) \sigma_z]
 %\end{split}
 \end{eqnarray}
 where $k_-= a (k_x-k_y)$, $k_+=a (k_x+k_y)$ and
 \begin{equation}
  f(J) \equiv 4\left(\frac{\pi - \frac{4JT}{5}}{\sin \left(\frac{4JT}{5}\right)}\right) \sin^2\left(\frac{JT}{5}\right) \cos\left(\frac{JT}{5}\right)
  \end{equation}
When $\frac{4JT}{5}=\pi$, the function $f(JT) \rightarrow 1$, and this has the structure of the 2D massless Dirac equation: $(H_{\rm{eff}}T-\pi) \approx a(k_x-k_y)\sigma_x+a(k_x+k_y)\sigma_y$. 

Boundaries, or spatial inhomogeneities can be accommodated in the real-space formalism of Eq.~(\ref{meq}), and can lead to edge modes \cite{kitagawa2010, rudner2013}.

In Fig.~\ref{bandimg}, we plot the band structure $\epsilon(k_y)$ for the system in a strip-geometry (open boundary conditions in the $x$-direction and periodic in the $y$-direction) for $JT=0.5 \pi, 1.0\pi, 2.0\pi$. We will later consider more realistic experimental geometries. The phase at $JT=0.5 \pi$ and $JT=1.0 \pi$ is topologically trivial and there are no edge modes in the system. However, as one would expect from Eq.~\ref{hsmallk}, at $JT=J_cT=\frac{5}{4} \pi$ the gap at $\epsilon=\pi/T$ closes. The gap then reopens for $J > J_c$, leaving edge modes connecting the top of the band to the bottom as seen in Fig.~\ref{bandimg}(c).
As argued by Rudner et. al. \cite{rudner2013} the topological invariant in this case is a ``winding number" $W$ calculated from the full time evolution operator $U(t,k_x, k_y)$:
\begin{equation} \label{windeq} 
\begin{split}
W[U]= \frac{1}{8\pi^2} & \int{dt dk_x dk_y} \\
                                   & \cdot Tr(\tilde{U}^{-1} \partial_t \tilde{U} \cdot [\tilde{U}^{-1} \partial_{k_x} \tilde{U}, \tilde{U}^{-1} \partial_{k_y} \tilde{U}]).\\
  \end{split}                                
  \end{equation}
 where 
 \begin{equation}
 \tilde{U}(t)= U(2t)\Theta(T/2-t)+U_{\rm{eff}}(2T-2t) \Theta(t-T/2) 
 \end{equation}
In the topologically trivial phase $W=0$, meaning one can continuously deform $U(t)$ into $U_{\rm{eff}}(t)\equiv e^{-i H_{\rm{eff}}t}$. In the topologically nontrivial phase (for instance at $JT=2.0 \pi$) $W=1$ and there is no continuous path between them. Interestingly, the Chern numbers for the bulk bands of $H_{\rm{eff}}$ are zero for all $J$ \cite{rudner2013}.

\section{Imaging Edge States} \label{secsim}

In this section we discuss an experimental method for imaging edge states in the Floquet model discussed in the previous section.

\subsection{Edge State Physics}

\begin{figure}[tbh]
\hbox{\hspace{-1.8em}
\includegraphics[width=0.5\textwidth]{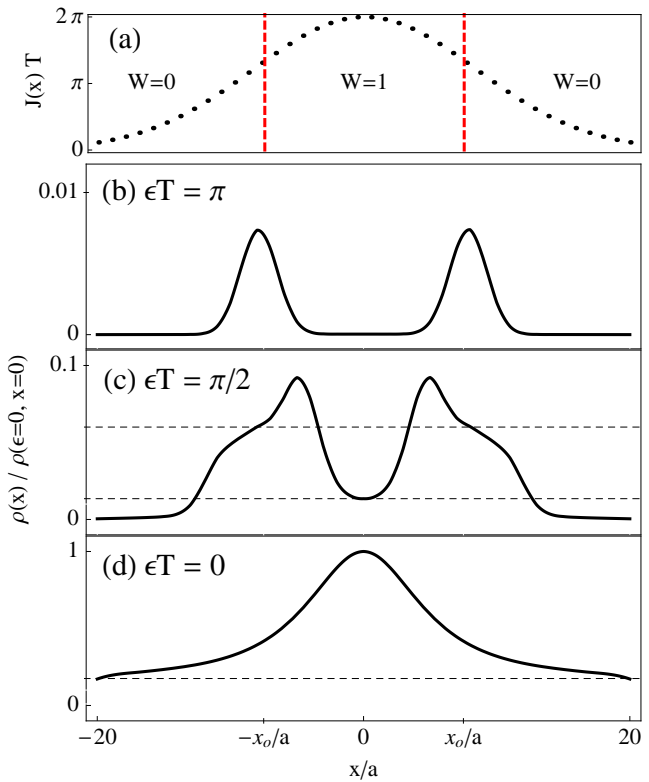}}
 \vspace{-0.0em}
\caption{(Color online) (a) Spatially dependent hopping  $J(x)$. (b,c,d) The corresponding local density of states $\rho(\epsilon, x)$ at $\epsilon T=\pi,\pi/2,0$ (see Eq.~\ref{ldoseq}). The red dashed lines in the top graph separate spatial regions that are in different topological phases. Mid-gap edge states at $\epsilon T=\pi$ are spatially localized at the boundary $\pm x_o$ between the different phases. Dashed horizontal lines are drawn to help visualize the strongest features.}
\label{ldosimg}
\end{figure}

Edge states can appear at the boundary between topologically distinct phases. In solid state models, this could be a boundary between ``vacuum" and a topological insulator or a domain boundary in a system with spatially modulated parameters~\cite{katan2013}. In the model considered here, the simplest interface to engineer is between the states in Fig.~\ref{bandimg}(a) and Fig.~\ref{bandimg}(c). As such we envision a spatially dependent hopping generated by the spatial profile of the laser beams creating the lattice (see Sec.~\ref{secexp}). Such spatially dependent hopping has precedence in cold atom experiments: Mathy et. al. recently proposed using a similar approach to help attain magnetic order in the Fermi-Hubbard model \cite{mathy2012}. Fig.~\ref{ldosimg}(a) shows a typical hopping profile $J(x)T= 2\pi \exp{[-\frac{(x-L/2)^2}{2\sigma^2}]}$. Also shown are the local values of the winding number $W$ as predicted by a local density approximation. Here we take periodic boundary conditions in the $y$-direction. This geometry is purely for theoretical convenience. In Sec.~\ref{secsim}B we consider the more experimentally relevant geometry where $J$ varies with $r=\sqrt{(x-\frac{L}{2})^2+(y-\frac{L}{2})^2}$. One expects an edge mode at the interface where $JT=J_cT =\frac{5}{4} \pi$. In our strip geometry we label this location as $x_o$. For the graph in Fig.~\ref{ldosimg}(a), $\sigma= \frac{L}{2\sqrt{6}}$,  $L=40a$, and $x_o \approx 7.5a$ where $L$ is the system size.

Fig.~\ref{ldosimg} (b) through (d) shows slices of the single-particle local density of states $\rho(\epsilon,x)$, 
\begin{equation} \label{ldoseq}
\rho(\epsilon,x)= \sum_n \int{dy |\psi_n(x,y)|^2} \delta(\epsilon-\epsilon_n)
\end{equation}
where $\psi_n$ and $\epsilon_n$ are the eigenstates and eigenenergies of the Floquet Hamiltonian $H_{\rm{eff}}$; we have broadened the delta function in Eq.~\ref{ldoseq} to a Lorentzian of width $0.05/T$. As expected, we find a density of mid gap edge states at the spatial locations where the hopping parameter $J(x,y)$ crosses between distinct topological regions. As is clear from Fig.~\ref{bandimg}, these mid-gap states should be visible at $\epsilon T=\pi$ (see Fig.~\ref{ldosimg}(b)). There are proposals to spectroscopically detect such edge states \cite{goldman2012, goldman2013eur}.

\begin{figure}
\hbox{\hspace{-0em}
\includegraphics[width=0.45\textwidth]{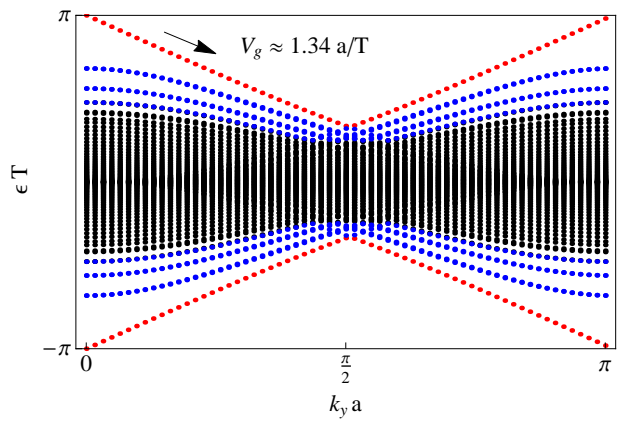}}
 \vspace{-0.0em}
\caption{(Color online) Dispersion relation $\epsilon(k_y)$ for a system with spatially dependent hopping $J(x)$ (see Fig.~\ref{ldosimg}). The red points specify the energies of ``massless" edge states with dispersion  $(\epsilon(k_y)- \frac{\pi}{T} ) \propto k_y$. These lie at the boundary between the $W=1$ and $W=0$ regions in Fig.~\ref{ldosimg}(a). The edge-state group velocity $V_g$ is given by slope of these lines $V_g = \frac{\partial \epsilon}{\partial k_y} \approx 1.34 a/T$. The blue points specify the energies of ``massive" states with dispersion ($\epsilon_n(k_y) - \frac{\pi}{T}) \propto \sqrt{\widetilde{m_n}^2+V_g^2 k_y^2}$ (see Eq.~\ref{edgediseq}).}
\label{shopdispimg}
\end{figure}

The structure of the states near the boundary is elucidated by expanding $H_{\rm{eff}}$ about $k_x=k_y=0$ (Eq.~\ref{hsmallk}) and $x=x_o$ .
Linearizing $J(x)$ at $x=x_o$ and squaring both sides of Eq.~\ref{hsmallk} gives 
\begin{equation}
(H_{\rm{eff}}T -\pi)^2 = (\frac{4T}{5} J'(x_o)(x-x_o))^2 + 2a^2 (k_x^2+k_y^2)
\end{equation}
In the strip geometry considered in this section, $k_y$ is a good quantum number while $k_x$ should be interpreted as a differential operator $k_x= \frac{1}{i} \partial_x$.
This is just an harmonic oscillator Hamiltonian in the $x$-direction plus a constant proportional to $k_y^2$. The energy spectrum of $H_{\rm{eff}}$ for the $\epsilon >0$  branch is then given by 

\begin{equation} \label{edgediseq} 
\epsilon_n (k_y)= \frac{\pi}{T}- \sqrt{m_n^2+ V_g^2 k_y^2}
\end{equation}
where $m_n= \sqrt{n \times 2 \sqrt{2} a (\frac{4T}{5} J'(x_o))}$, $V_g= \sqrt{2} \frac{a}{T}$, and $n \geq 0$ is an integer. There is one linearly dispersing ``massless" edge mode ($n=0$) and ladder of ``massive" modes localized near $x_o$ with effective mass $m_n \propto \sqrt{n}$. These analytic results match the numerical results shown in Fig.~\ref{shopdispimg} (red and blue points corresponding to $n=0$ and $4 \geq n>0$, respectively) within $\sim 10 \%$ error. The approximations we have used here improve for smaller $J'(x_o)$.

\subsection{Imaging Protocol}

\begin{figure} \vspace{-0.0em}
\hbox{\hspace{-2.0em}
\includegraphics[width=0.5\textwidth]{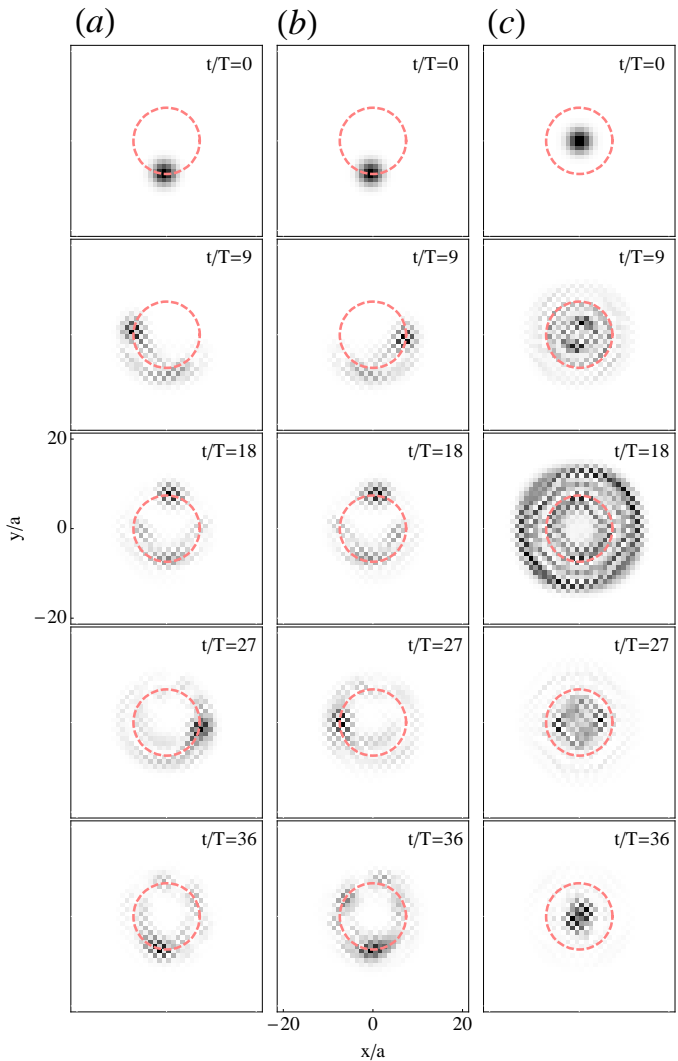}}
\caption{(Color online) Real space density showing chiral edge modes. Darker colors represent higher density. Column (a) shows the dynamics resulting from the experimental protocol described in Sec.~\ref{secsim}-B with spatially dependent hopping $J(x,y)T= 2\pi \exp{[-12\frac{(x-L/2)^2+(y-L/2)^2}{L^2}]}$ where $L=40 a$.  At time $t=0$ a condensate is placed in the lattice. A wavepacket moves in the clockwise direction along the circular boundary separating two topologically distinct regions (dashed red line). Column (b) shows the dynamics resulting from running the steps shown in Fig.~\ref{modelimg} in reverse. The edge state wave packet now moves in the counter-clockwise direction. Column (c) shows the dynamics resulting after initializing a wave packet away from the circular boundary. In this case there is no observable edge-state propagation.}
\label{edgeimg}
\end{figure}

To experimentally observe the edge states in this setup, we suggest watching the motion of a wave-packet of bosons. Fig.~\ref{edgeimg} shows the time evolution of the boson density in a simulation with spatial dependent hopping given by $J(x,y)T= 2\pi \exp{[-\frac{(x-L/2)^2+(y-L/2)^2}{2\sigma^2}]}$. When the system is initialized with a gaussian wave packet localized at the boundary, we find a clearly identifiable wave packet propagating in the clockwise-direction along a circle of radius $7.5 a$ (Fig.~\ref{edgeimg}(a)). The time it takes for the wave packet to propagate around the circle once, $t' \approx 35 T$, is consistent with the group velocity $V_g$ we calculated in Fig.~\ref{shopdispimg}: $t' = \frac{\rm{circumference}}{V_g} \approx \frac{2\pi \times 7.5 a}{1.34 a/T}\approx 35 T$. The packet does spread somewhat, as several modes are occupied. 

Fig.~\ref{edgeimg}(b) shows the result of running the steps of the model (Fig.~\ref{modelimg}) in reverse. Here one finds that the edge state wave packet moves with the opposite chirality. By contrast, if the packet is initialized away from the boundary (Fig.~\ref{edgeimg}(c)) it undergoes dynamics in which it sequentially expands and contracts.

For these calculations we used a large grid extending in both the $x$ and $y$ directions with open boundary conditions. We verified that the edge of our grid did not affect our results.

%As a test, we also ran simulations with a spatially constant hopping $J=2\pi/T$ (see Fig (FIGURE WITH CONSTANT J)); in this case all we find is diffusion of the initial wave packet with no identifiable edge dynamics. %

\section{Experimental Implementation} \label{secexp}
One can implement this model experimentally using an optical lattice driven with time-dependent phases and amplitudes. In particular, we envision a potential of the form 
\begin{equation} \label{poteq}
\begin{split}
&V(x,y,t)=  A (\sin^2(k_{L_1}x) + \sin^2 (k_{L_1}y)) \\
& \begin{split} - C(x,y,t) \times & (\sin^2[\frac{k_{L_1}}{2}(x+y+\phi_1(t))] \\
                       &+\sin^2[\frac{k_{L_1}}{2}(x-y+\phi_2(t))]) \\
                       \end{split} \\
\end{split}
\end{equation}
where $C(x,y,t)=C(x,y)$ for steps $1-4$ and $C(x,y,t)=0$ for step 5. This fifth step can be omitted, if desired. $\phi_1(t)=+\pi/2$ for steps 1 and 2 and $\phi_1(t)=-\pi/2$ for steps 3 and 4;  $\phi_2(t)=+\pi/2$ for steps 1 and 4 and $\phi_2(t)=-\pi/2$ for steps 2 and 3. This potential is illustrated in Fig.~\ref{modelimg} for spatially uniform $C$. If $A$ is sufficiently large, hopping only occurs between neighboring pairs of sites. The spatially dependent $C(x,y,t)$--which generically decreases away from the center as described below-- implies that the hopping is stronger at the center than the edge as in Fig.~\ref{ldosimg}(a). Additionally, a deep lattice along the $z$-direction restricts motion to two dimensions.

One can create the first term of Eq.~\ref{poteq} with two independent sets of counter propagating lasers. The second term in Eq.~\ref{poteq} is created with two sets of \textit{red-detuned} lasers with wave-vectors $\vec{k}_{L_2}= (\frac{k_{L_1}}{2}, \pm \frac{k_{L_1}}{2}, q)$. The resulting potential in the $x$-$y$ plane does not depend on $q$, but allowing such a term gives additional design flexibility. By modulating the amplitudes and phases of the lasers one would control the the time-dependence of $C$, $\phi_1$, and $\phi_2$. The finite beam waists of these lasers naturally yield a profile $C(x,y) \approx \exp(-(x^2+y^2)/2\sigma^2)$. With this spatial dependence the barrier between neighboring lattice points is maximally reduced at $(x,y)=(0, 0)$ but grows as $x$ and $y$ increase, similar to Fig.~\ref{ldosimg}(a). The spatial variation of the hopping parameter can further be controlled by changing the profile of the laser \cite{gotlibovych2014}.

\section{Conclusions}
We have proposed an experiment that realizes a Floquet topological insulator in an optical lattice, and we have demonstrated an experimental protocol that allows for the direct observation of topologically protected edge states. Using numerical simulations, we have shown that by imaging the evolution of a wavepacket, chiral edge states can be observed at the boundary between two distinct topological phases. Our proposal provides a simple and direct way to realize and probe topologically ordered quantum states using ultracold atoms.

\section{Acknowledgements}
This material is based upon work supported by the National Science Foundation Graduate Research Fellowship under Grant No. DGE-1144153 as well as work supported by the National Science Foundation under Grant No. PHY-1068165.

\bibliography{Floquet}

\end{document}